\documentclass[useAMS,usenatbib]{mn2e}
\usepackage{mathrsfs}
\usepackage{graphicx,amssym,color}
\newcommand{\beq}{\begin{equation}}
\newcommand{\eeq}{\end{equation}}
\newcommand{\beqa}{\begin{eqnarray}}
\newcommand{\eeqa}{\end{eqnarray}}

\usepackage{color}
\def\gsim { \lower .75ex \hbox{$\sim$} \llap{\raise .27ex \hbox{$>$}} }
\def\lsim { \lower .75ex \hbox{$\sim$} \llap{\raise .27ex \hbox{$<$}} }

\newcommand{\bc}{\begin{center}}
\newcommand{\ec}{\end{center}}

\title[Missing Massive Satellites around the Milky Way]
      {The Missing Massive Satellites of the Milky Way}
\author[Jie Wang et al. ]
       { Jie Wang$^{1}$\thanks{Email: jie.wang@durham.ac.uk}, Carlos
         Frenk$^1$, Julio F. Navarro$^2$, Liang Gao$^3$ and Till Sawala$^1$
  \\
$^1$Institute for Computational Cosmology, Department of Physics,
  University of Durham, South Road, Durham, DH1 3LE, UK\\
$^2$Department of Physics and Astronomy, University of Victoria, PO
Box 3055 STN CSC, Victoria, BC, V8W 3P6 Canada\\
$^3$National Astronomical Observatories, Chinese Academy of Science,
Beijing, 100012, China\\
}

\begin{document}

\date{\today}

\pagerange{\pageref{firstpage}--\pageref{lastpage}}
\pubyear{2012}

\maketitle

\label{firstpage}

\begin{abstract}
  Recent studies suggest that only three of the twelve brightest
  satellites of the Milky Way (MW) inhabit dark matter halos with
  maximum circular velocity, $V_{\rm max}$, exceeding $\sim 30$
  km/s. This is in apparent contradiction with the $\Lambda$CDM
  simulations of the Aquarius Project, which suggest that MW-sized
  halos should have at least $8$ subhalos with $V_{\rm max}>30$ km/s.
  The absence of luminous satellites in such massive subhalos is thus
  puzzling and may present a challenge to the $\Lambda$CDM
  paradigm. We note, however, that the number of massive subhalos
  depends sensitively on the (poorly-known) virial mass of the Milky
  Way, and that their scarcity makes estimates of their abundance from
  a small simulation set like Aquarius uncertain. We use the
  Millennium Simulation series and the invariance of the scaled
  subhalo velocity function (i.e., the number of subhalos as a
  function of $\nu$, the ratio of subhalo $V_{\rm max}$ to host halo
  virial velocity, $V_{200}$) to secure improved estimates of the
  abundance of rare massive subsystems. In the range $0.1<\nu<0.5$,
  $N_{\rm sub}(>\nu)$ is approximately Poisson-distributed about an
  average given by $\langle N_{\rm sub} \rangle=10.2\,
  (\nu/0.15)^{-3.11}$. This is slightly lower than in Aquarius halos,
  but consistent with recent results from the Phoenix Project. The
  probability that a $\Lambda$CDM halo has $3$ or fewer subhalos with
  $V_{\rm max}$ above some threshold value, $V_{\rm th}$, is then
  straightforward to compute. It decreases steeply both with
  decreasing $V_{\rm th}$ and with increasing halo mass. For $V_{\rm
    th}=30$~km/s, $\sim 40\%$ of $M_{\rm halo}=10^{12} \, M_\odot$
  halos pass the test; fewer than $\sim 5\%$ do so for $M_{\rm
    halo}\gsim \, 2\times 10^{12}\, M_\odot$; and the probability
  effectively vanishes for $M_{\rm halo}\gsim \, 3\times 10^{12}\,
  M_\odot$. Rather than a failure of $\Lambda$CDM, the absence of
  massive subhalos might simply indicate that the Milky Way is less
  massive than is commonly thought.
\end{abstract}

\begin{keywords}
\end{keywords}

\section{Introduction}
\label{SecIntro}

The striking difference between the relatively flat faint-end slope of
the galaxy stellar mass function and the much steeper cold dark matter
halo mass function is usually reconciled by assuming that the
efficiency of galaxy formation drops sharply with decreasing halo mass
\citep[see, e.g.,][]{white1991}.  Semi-analytic models of
galaxy formation have used this result to explain the relatively small
number of luminous satellites in the Milky Way (MW) halo, where
$\Lambda$CDM simulations predict the existence of thousands of
subhalos massive enough, in principle, to host dwarf galaxies.  In
these models, the small number of MW satellites reflects the
relatively small number of subhalos massive enough to host luminous
galaxies \citep[see,
e.g.,][]{Kauffmann1993,Bullock2000,Benson2002,Somerville2002,Cooper2010,Li2010,Maccio2010,Guo2011a,Font2011a}.

This is a model prediction that can be readily tested observationally,
given the availability of radial velocity measurements for hundreds of
stars in the dwarf spheroidal satellites of the Milky Way. Combined with
photometric data, radial velocities tightly constrain the total mass
enclosed within the luminous radius of these satellites
\citep{Walker2009,Wolf2010}. The latter correlates strongly with the
total dark mass of the dwarf, which is usually expressed in terms of
its maximum circular velocity, $V_{\rm max}$, a quantity less affected
than mass by tidal stripping \citep{Penarrubia2008b}.

Kinematical analyses of the Milky Way dwarf spheroidals have been
attempted by several authors in recent years, with broad consensus on
the results, at least for the best-studied nine brightest dwarf
spheroidal MW companions: Draco, Ursa Minor, Fornax, Sculptor, Carina,
Leo I, Leo II, Canis Venatici I, and Sextans
\citep[see, e.g.,][]{Penarrubia2008a,Strigari2008,Lokas2009,
Walker2009,Wolf2010,Strigari2010}. These studies suggest that some of
these galaxies may inhabit halos with $V_{\rm max}$ as low as $12$
km/s, and agree that all\footnote{One possible exception is Draco,
where the data might allow a more massive halo.} appear to inhabit
halos with values of $V_{\rm max}$ below a low threshold, $V_{\rm th}
\sim 30$~km/s. Only three dwarf irregular satellites -- the
Magellanic Clouds and the Sagittarius dwarf -- may, in principle,
inhabit halos exceeding this threshold.

The most straightforward interpretation of this result is that massive
subhalos in the Milky Way are rare. However, as argued recently by
\citet{Boylan-Kolchin2011a,Boylan-Kolchin2011b}, this is at odds with
the results of the Aquarius Project, a series of N-body simulations of
six different halos of virial\footnote{Unless otherwise noted, we
  define virial quantities as those corresponding to spheres that
  enclose a mean overdensity $\Delta=200$ times the critical density
  for closure. $M_{200}$, for example, corresponds to the mass within
  the virial radius, $r_{200}$. When other values of $\Delta$ are
  assumed the subscript is adjusted accordingly.} mass in the range
$0.8 < M_{200}/10^{12}\, M_\odot < 1.8$. \cite{Boylan-Kolchin2011a}
noted that the largest subhalos in these simulations are significantly
denser than inferred for the halos that host the brightest dwarf
spheroidals in the Milky Way.

As discussed by \cite{Parry2011} and \cite{Boylan-Kolchin2011b}, the
discrepancy can be traced to the fact that the largest Aquarius
subhalos are significantly more massive or, equivalently, have too
large a value of $V_{\rm max}$ to be compatible with the measured
kinematics of the brightest dwarf spheroidals. Specifically, the
Aquarius halos have, on average, $\sim 8$ subhalos with $V_{\rm
  max}>30$ km/s within the virial radius, larger than the $V_{\rm
  max}$ of the brightest dwarf spheroidals, prompting questions about
why these massive subhalos fail to host luminous satellites in the
Milky Way. If this result holds, it may point to a failure of our
basic understanding of how galaxies populate low mass halos or, more
worryingly, of the $\Lambda$CDM paradigm itself.

Two issues may affect these conclusions. One is that the Aquarius
Project simulation set contains only $6$ halos and, therefore,
estimates of the abundance of rare massive subhalos are subject to
substantial uncertainty. The second point is that the number of
massive subhalos is expected to depend sensitively on
the virial mass of the host halo, which is only known to within a
factor of $2$-$3$ for the Milky Way.

We address these issues here by using large numbers of well-resolved
halos identified in the Millennium Simulation series
\citep{Springel2005a,Boylan-Kolchin2009}. This is possible because, in
agreement with earlier work, we find that the abundance of subhalos,
when scaled appropriately, is independent of halo mass \citep[see,
e.g.,][]{Moore1999b,Kravtsov2004,Springel2008b}. We use this to derive
improved estimates of the average number of massive subhalos, as well
as its statistical distribution. The probability that a halo has as
few massive subhalos as the Milky Way can then be evaluated, both as a
function of host halo mass and/or subhalo mass threshold.

This paper is organized as follows. Sec.~\ref{SecMS} describes briefly
the simulations we used in our analysis. We present our main results
in Sec.~\ref{SecRes}, and end with a brief summary in
Sec.~\ref{SecConc}.

\begin{figure}
\bc
\hspace{-1.cm}
\resizebox{9cm}{!}
{\includegraphics{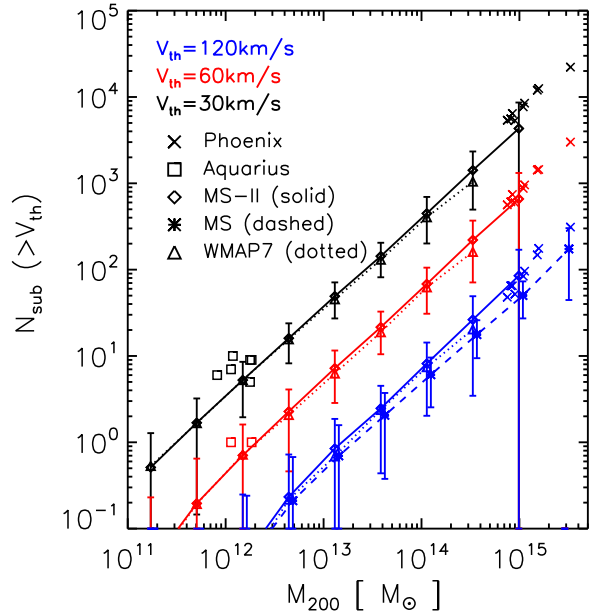}}\\%
\caption{The number of subhaloes with $V_{\rm max} \geq V_{\rm th}$ as
  a function of the virial mass of their host haloes, $M_{\rm 200}$,
  in the Millennium Simulations (MS and MS-II), as well as in the
  level-2 runs of the Aquarius and Phoenix Projects. Subhalos are
  identified within the virial radius, $r_{\rm 200}$, of their host
  systems. Different symbols correspond to each simulation, as
  labelled, and are coloured according to the value of the threshold,
  $V_{\rm th}$. Error bars denote the rms plus Poisson error in each
  mass bin. Note the nearly linear dependence of the number of
  subhalos with halo mass. Due to numerical resolution, few subhalos
  with velocities less than $\sim 100$ km/s are found in the MS
  simulation, so the $V_{\rm th}=30$ and $60$ km/s MS curves in this
  case are omitted for
  clarity. For massive, well-resolved halos the results are much less
  affected by numerical limitations and there is good agreement
  between MS and MS-II. Subhalo abundance is insensitive to small variations
  in the cosmological parameters. Triangles connected by a dotted line show
  results corresponding to a run that adopted the latest WMAP7
  parameters \citep{Komatsu2011}; in contrast, the Millennium
  Simulations adopted parameters consistent with the 1st-year analysis
  of WMAP data. } \label{FigMNv} \ec
\end{figure}

\begin{figure}
\bc
\hspace{-1.cm}
\resizebox{9cm}{!}{\includegraphics{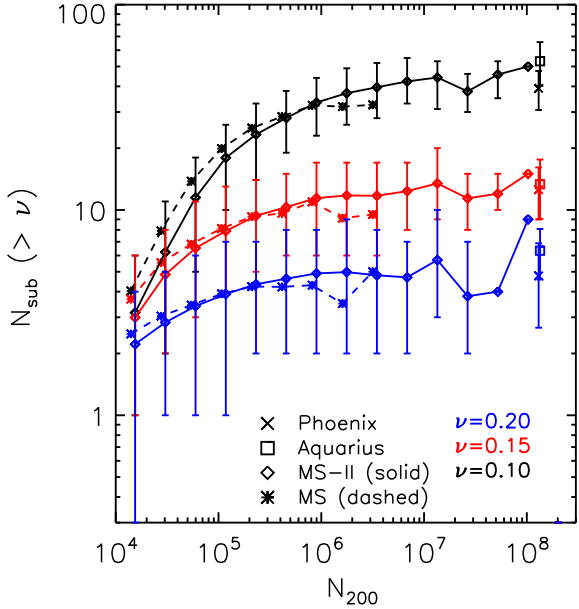}}\\%
\caption{The number of subhalos with maximum circular velocity
  exceeding a certain fraction, $\nu$, of their host halo virial
  velocity, as a function of $N_{200}$, the number of particles within
  the virial radius of the host. Curves for three values of $\nu=0.1$,
  $0.15$, and $0.2$ are shown. Error bars for MS and MS-II indicate
  the 10 and 90 percentile in each bin, and are omitted when the bin
  contains a single halo. The excellent agreement between results
for MS and MS-II at given $N_{200}$ reflects the halo mass invariance
of the $N_{\rm sub}(>\nu)$ function; each particle is $125\times$ more
massive in MS than in MS-II. Results converge for well-resolved halos
(i.e., those with large $N_{200}$). As expected, the smaller $\nu$ the
larger the minimum number of particles, $N_{200}^{\rm min}$, needed to
obtain converged results.}
\label{FigNN200}
\ec
\end{figure}

\section{Simulations}
\label{SecMS}

The two Millennium simulations (MS; \citealt{Springel2005a} and MS-II; \citealt{Boylan-Kolchin2009})
provide the main datasets used in this study. Both are simulations of a flat WMAP-1
$\Lambda$CDM cosmogony with the following parameters: $\Omega_{\rm
  M}=0.25$, $\Omega_b=0.045$, $h=0.73$, $n_s=1$ and
$\sigma_8=0.9$. 

The MS run evolved a box $500$ Mpc/$h$ on a side, with $2160^{3}$
particles of mass $m_p=8.6 \times 10^8 M_{\odot}/h$. MS-II evolved the
same total number of particles in a box $1/125$ the volume of MS and
had, therefore, 125 better mass resolution ($m_p=6.885 \times 10^6
M_{\odot}/h$). The nominal spatial resolution is given by the
Plummer-equivalent gravitational softening, which is $\epsilon_P=5
$ kpc$/h$, and $1$ kpc$/h$ for the MS and MS-II runs, respectively.

We also use halos from the Aquarius Project \citep{Springel2008b} and
the Phoenix Project \citep{Gao2012} (level-2 resolution). These are
ultra high-resolution simulations of six MW-sized halos ($M_{200} \sim
10^{12}M_{\odot}$) and nine cluster-sized halos ($M_{200}\sim 10^{15}
M_\odot$), each resolved with a few hundred million particles within
the virial radius.  

The normalization of the power spectrum adopted in these simulations
is slightly higher than favoured by the latest WMAP dataset
\citep[WMAP7;][]{Komatsu2011}, but this is expected to affect the
abundance of halos of given virial mass rather than the mass function
of subhalos, which is the main focus of our study. We have verified
this explicitly by analyzing a $1620^3$-particle simulation of a
$70.4$ Mpc/$h$ box that adopts the WMAP7 cosmological parameters (see
Fig.~\ref{FigMNv}). The particle mass in this run is $6.20 \times
10^6 \, M_\odot/h$ and gravitational interactions were softened with
$\epsilon_P=1$ kpc/$h$.

Halos and subhalos are identified in all simulations by {\sc subfind}
\citep{Springel2001a}, a recursive algorithm that identifies
self-bound structures and substructures in N-body simulations.

\begin{figure}
  \bc \hspace{-1.cm}
  \resizebox{9cm}{!}{\includegraphics{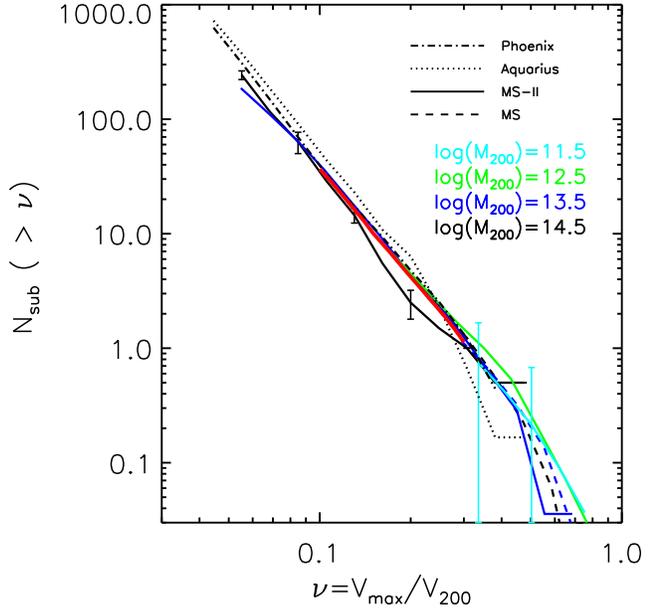}}\\%
  \caption{The scaled subhalo velocity function, i.e., the number of
    subhalos with maximum circular velocity exceeding a certain
    fraction, $\nu=V_{\rm max}/V_{200}$, of the host halo virial
    velocity. Dotted and dot-dashed curves show averages for the six
    Aquarius halos and nine Phoenix halos, respectively. Dashed and
    solid curves correspond to MS and MS-II. Four curves are shown for
    each, corresponding to averages over all halos in mass bins of
    width $0.1$ dex centred at $\log_{10}M_{200}/M_\odot=11.5$,
    $12.5$, $13.5$, and $14.5$. Error bars (shown only for the lowest
    and highest 
    mass bins) indicate the {\em rms} scatter in each bin. Only halos
    satisfying the constraint $N_{200}>N_{200}^{\rm min}(\nu)$ are
    used.  All simulations are in good agreement when well-resolved
    halos are considered. The scaled subhalo velocity function is thus
    nearly invariant with mass. See text for further
    discussion.}  \label{FigNnu} \ec
\end{figure}

\section{Results}
\label{SecRes}

We first investigate the scale invariance and other statistical
properties of the distribution of subhalo $V_{\rm max}$ and then apply
our results to subhalos in the Milky Way.

\subsection{Subhalo  $V_{\rm max}$ distribution}
\label{SecRes1}

Fig.~\ref{FigMNv} shows, as a function of host halo virial mass, the
total number\footnote{Unless otherwise noted, we identify subhalos
  within the virial radius, $r_{200}$, of the host halo.} of subhalos
with maximum circular velocity, $V_{\rm max}$, exceeding a specified
velocity threshold, $V_{\rm th}$. Results are shown for three
different values of $V_{\rm th}$. The average number of subhalos in
each halo mass bin is shown by symbols connected by solid (MS-II) or
dashed (MS) lines. Individual level-2 Phoenix and Aquarius halos are
shown by crosses and open squares, respectively. WMAP7 results are
shown by open triangles connected by a dotted line.

Fig.~\ref{FigMNv} illustrates that: (i) the number of subhalos depends
roughly linearly on halo mass and increases strongly with decreasing
velocity threshold, and that (ii) the slight change in cosmological
parameters from WMAP1 to WMAP7 has a negligible effect on subhalo
abundance. 

Fig.~\ref{FigMNv} also shows that numerical resolution limits the halo
mass and velocity threshold for which convergence in subhalo
abundances is achieved. Indeed, there are fewer subhalos in MS, the
simulation with poorest mass resolution; so few with velocities less
than $\sim 100$ km/s that the $V_{\rm th}=30$ and $60$ km/s MS curves
have been omitted for clarity.  When halos and subhalos are resolved
with enough particles, however, the results converge well. For $V_{\rm
  th}=120$ km/s, MS, MS-II, and WMAP7 halos yield similar numbers of
subhalos over the whole halo mass range considered, despite the fact
that, at given $M_{200}$, MS-II and WMAP7 halos have $\sim 125\times$
more particles than their MS counterparts.  Furthermore, the results
for Phoenix and Aquarius are in good agreement with MS-II, even though
halos in MS-II have $700\times$ fewer particles than Aquarius and
$2\times$ fewer particles than Phoenix, respectively.

We explore the requirements for numerical convergence in more detail
in Fig.~\ref{FigNN200}, where we plot, as a function of the total
number of particles within the virial radius, $N_{200}$, the average
number of subhalos with $V_{\rm max}$ exceeding a certain fraction of
the host halo virial velocity: $N_{\rm sub}(>\nu)$, for $\nu=V_{\rm
  max}/V_{200}=0.10$, $0.15$, and $0.20$. Results are shown for MS and
MS-II halos with dashed and solid curves, respectively.

This figure highlights two important points. One is that at given
$\nu$ there is good agreement between all simulations provided that
halos are resolved with enough particles. The second is that, when the
first condition is met, $N_{\rm sub}(>\nu)$ is independent of halo
mass. (Recall that, at fixed $N_{200}$, MS halos are $125\times$ more
massive than their MS-II counterparts.) This agreement, together with
the fact that the $N_{\rm sub}(>\nu)$ curves plateau at large values
of $N_{200}$, imply that the {\it scaled} subhalo velocity function
(i.e., the number of subhalos as a function of $\nu=V_{\rm
  max}/V_{200}$) is invariant over many decades in halo
mass. Fig.~\ref{FigNN200} also makes clear that numerical convergence
requires that a halo be resolved with a total number of particles
above some ($\nu$-dependent) minimum number, $N_{200}^{\rm min}$
(listed in Table~1). The
converged values agree well, within the statistical uncertainty,
with the results for Phoenix and Aquarius halos.

We show the invariance of subhalo abundance explicitly in
Fig.~\ref{FigNnu}, where $N_{\rm sub}(>\nu)$ is plotted for MS and
MS-II halos grouped in 4 bins of different halo mass. Only halos
satisfying the $N_{200}>N_{200}^{\rm min}$ constraint are used here.
These results confirm earlier work \citep[see,
e.g.,][]{Moore1999b,Kravtsov2004,Zheng2005,Weinberg2008,Springel2008b},
and imply that we can combine all well-resolved halos into one large
sample to derive robust estimates of the statistical {\it
  distribution} of $N_{\rm sub} (>\nu)$. 

\begin{figure}
\bc
\hspace{0.1cm}
\resizebox{8cm}{!}{\includegraphics{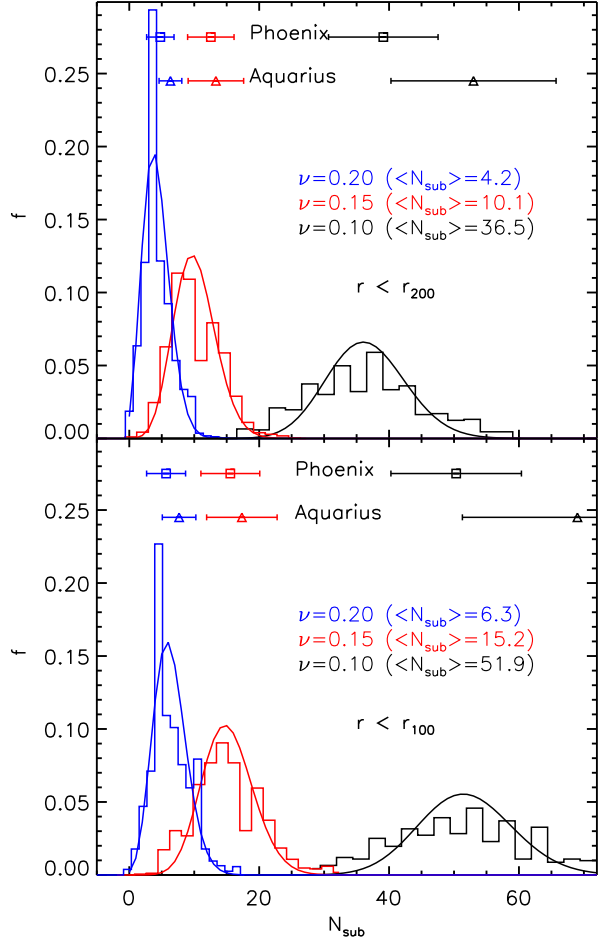}}\\%
\caption{The distribution of $N_{\rm sub}(>\nu)$ for $\nu=0.1$, $0.15$, and
  $0.2$, computed for well-resolved MS and MS-II halos; i.e., those
  with particle numbers exceeding $N_{200}^{\rm min}(\nu)$ (as given
  in Table~\ref{TabN200Min}). The top panel refers to all subhalos
  within the virial radius, $r_{200}$; the bottom panel to subhalos
  within a radius roughly $30\%$ larger, $r_{100}$.  The average and rms for Aquarius and
  Phoenix halos are shown at the top of the plot. Note that subhalos in
  Aquarius seem slightly overabundant relative to either Phoenix
  or the Millennium Simulations, but still well within the statistical
  uncertainty. The $N_{\rm sub}(>\nu)$ distribution is well approximated by a
  Poisson distribution: the solid curves show Poisson distributions
  with the same averages as each histogram.}
\label{FigDistNnu}
\ec
\end{figure}

This is shown in Fig.~\ref{FigDistNnu} for $\nu=0.1$, $0.15$, and
$0.2$, computed using all MS and MS-II halos with
$N_{200}>N_{200}^{\rm min}(\nu)$, as given in
Table~\ref{TabN200Min}. In the top panel, which corresponds to
subhalos identified within $r_{200}$, the histograms 
show the $N_{\rm sub}(>\nu)$ distributions for
the $614$, $3070$, and $6867$ halos that satisfy, respectively, the
minimum particle number constraint.  The bottom panel shows subhalo
numbers identified within a slightly larger radius, $r_{100}$, which
is on average $\sim 30\%$ larger than $r_{200}$.  (The mean and rms
dispersion of the distributions of $N_{\rm sub}(>\nu)$ are listed in Table~1.) Note that the
results obtained for MS and MS-II are in excellent agreement with the
results obtained for Phoenix halos. Subhalos in Aquarius are slightly
overabundant, perhaps because of small biases in their assembly
histories \citep{Boylan-Kolchin2010}, but still consistent with the MS
and MS-II results given the large variance ($\sigma_{N_{\rm sub}}^2$) expected from
the sample of only 6 Aquarius halos. Note as well that, as expected,
the larger volume encompassed by $r_{100}$ yields larger subhalo
numbers than found when identifying subhalos only within $r_{200}$.
For $r<r_{200}$, the average $N_{\rm sub}(>\nu)$ is a steep function
of $\nu$, well approximated, in the range $0.1<\nu<0.5$, by
\begin{equation}
 \langle N_{\rm sub}\rangle (>\nu)=10.2\, (\nu/0.15)^{-3.11}.\label{EqNnu}
\end{equation}
Fig.~\ref{FigDistNnu} also shows that the distribution of $N_{\rm
  sub}(>\nu)$ follows Poisson statistics closely; the solid curves are
actually {\it not} fits, but just Poisson distributions with the same
average as each of the histograms.  Clearly, these provide a good
description of the distribution of $N_{\rm sub}(>\nu)$ at fixed
$\nu$. This conclusion is supported by earlier work \citep[see,
e.g.,][]{Kravtsov2004,Boylan-Kolchin2010}, as well as by the data
listed in Table~\ref{TabN200Min}: the average number of subhalos is
roughly similar to the variance, as expected from a Poisson process.

\begin{figure}
\bc
\hspace{-1.cm}
\resizebox{9cm}{!}{\includegraphics{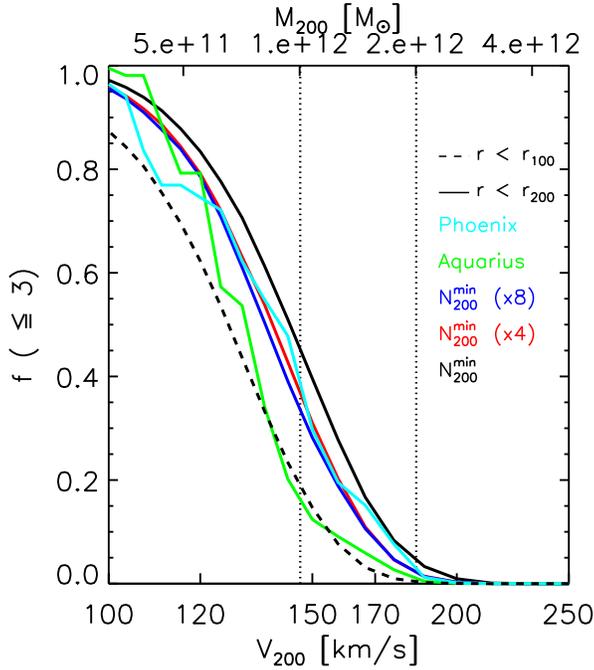}}\\%
\caption{Probability that a halo contains $3$ or fewer subhalos with
  $V_{\rm max}>30$ km/s, as a function of halo mass (top tickmarks) or
  virial velocity (bottom tickmarks). The solid black curve
  corresponds to assuming Poisson statistics and that $N_{\rm
    sub}(>\nu)=10.2\, (\nu/0.15)^{-3.11}$, the average number of
  subhalos within the virial radius of well-resolved  MS and MS-II halos with particle
  numbers exceeding $N_{200}^{\rm min}$ (see
  Table~\ref{TabN200Min}). The sensitivity of the result to the
  assumed minimum number of particles is shown by the red and blue
  curves, which correspond to increasing the values of $N_{200}^{\rm
    min}(\nu)$ by factors of $4$ or $8$, respectively. Results using
  only the nine Phoenix or six Aquarius halos are shown in cyan and
  green, respectively. Note that because subhalos are slightly
  overabundant in Aquarius (see Fig.~\ref{FigDistNnu}) the
  probabilities are systematically lower than when considering either
  Phoenix halos or the Millennium simulations. The same is true if
  subhalos are identified within a radius larger than the virial
  radius. The dashed curve shows probabilities when the search radius
  around each halo is increased by roughly $30\%$ to $r_{100}$. }
\label{FigfV200}
\ec
\end{figure}

\subsection{Massive satellites in the Milky Way}
\label{SecRes2}

We can use these results to address the Milky Way missing massive
satellites problem highlighted in Sec.~\ref{SecIntro}. In particular,
it is straightforward to compute the probability that a halo has $X$
or fewer subhalos with $V_{\rm max}>30$ km/s within its virial radius,
once a virial mass (or, equivalently, a virial velocity, $V_{200}$)
has been assumed for the Milky Way. This is given by,
\begin{equation}
  f(\leq X)= \sum_{k=0}^X {\frac{\lambda_\nu^k}{k!}} e^{-\lambda_\nu}, \label{EqPoisson}
\end{equation}
where $\lambda_\nu=\langle N_{\rm sub}\rangle (>\nu)$ is given by
Eqn.~\ref{EqNnu}.

The solid black curve in Fig.~\ref{FigfV200} shows $f(\leq 3)$ as a
function of virial mass (upper tickmarks on the abscissa) or virial
velocity (lower tickmarks). The probability is a steep function of
the assumed halo mass: more that $40\%$ of $10^{12} M_\odot$ halos
pass this test, but only $\sim 5\%$ of $2\times 10^{12} M_\odot$
systems do so. The probability becomes negligible for $M_{200}\gsim
\, 3\times 10^{12} M_\odot$.  This suggests that the scarcity of
massive subhalos is best thought of as placing a strong upper
limit on the virial mass of the Milky Way, rather than as a failure
of the $\Lambda$CDM scenario.


It is important to assess the sensitivity of this conclusion to the
parameters assumed in this study. For example, should the velocity
threshold be placed at $25$~km/s, rather than at $30$~km/s, as argued by
\citet[][]{Boylan-Kolchin2011b}, the upper limit on the mass of the
Milky Way would become even more restrictive. The results, however,
could still be read from Fig.~\ref{FigfV200}, after shifting the
tickmarks by $30/25=1.2$ in the velocity axis or by $1.2^3=1.73$ in
the mass axis. Thus, for $V_{\rm th}=25$~km/s, a
probability of more than 5\% requires a halo mass $M_{200}< \, 1\times
10^{12} M_\odot$, rather than the $M_{200}< \, 2\times
10^{12} M_\odot$ appropriate to $V_{\rm th}=30$~km/s.


We have also examined the dependence of our results on $N_{200}^{\rm
  min}(\nu)$, the assumed minimum number of particles needed for
convergence (listed in Table~\ref{TabN200Min}). This is shown by the
red and blue curves in Fig.~\ref{FigfV200}, which correspond to
increasing $N_{200}^{\rm min}$ by a factor of $4$ and $8$,
respectively, before deriving $\langle N_{\rm sub} \rangle
(>\nu)$. Fig. ~\ref{FigfV200} makes clear that our results are quite
insensitive to such changes in $N_{200}^{\rm min}$ .

Since Phoenix halos have subhalo abundances in good agreement with
those in Eqn.~\ref{EqNnu}, our results would not change had we chosen
the nine Phoenix halos to compute $\langle N_{\rm sub} \rangle (>\nu)$
(see cyan curve in Fig.~\ref{FigfV200}). On the other hand, had we
chosen to derive $\langle N_{\rm sub}\rangle (>\nu)$ solely from the
six Aquarius halos, the slight overabundance of subhalos in these
systems would lead to stricter upper limits on the Milky Way halo
mass, as shown by the green curve in Fig.~\ref{FigfV200}. This result,
together with the fact that the average Aquarius halo mass ($1.42
\times 10^{12} \, M_\odot$) is uncomfortably close to the upper limit
discussed above is apparently the reason why
\citet{Boylan-Kolchin2011a} originally found such a strong discrepancy
between the Aquarius simulations and the Milky Way.

Finally, we need to consider the dependence of the number of subhalos
on the maximum radius used to identify substructure. The results
discussed above
refer to subhalos identified within the virial radius, $r_{200}$,
which is $\sim 200$ kpc for a $M_{200}=10^{12}\,M_\odot$ halo. This is
smaller than the maximum distance commonly adopted to identify dwarf
galaxies as Milky Way satellites; for example, Leo I is at roughly
$250$ kpc from the centre of the Galaxy. Therefore, it might be argued
that subhalos should be counted within a larger radius in order to
make a meaningful comparison. As shown in the bottom panel of
Fig.~\ref{FigDistNnu}, subhalos are roughly $\sim 50\%$ more abundant
within $r_{100}$ than within $r_{200}$. For a
$M_{200}=10^{12}\,M_\odot$ halo, $r_{100}\approx 270$ kpc, comparable
to the Galactocentric distance of Leo I.  In analogy
with Eqn.~1,  the average number of subhalos, $N_{\rm sub}(>\nu)$, 
within  $r_{100}$ is well approximated, in the range $0.1<\nu<0.5$, by
\begin{equation}
 \langle N_{\rm sub}\rangle (>\nu)=15.03\, (\nu/0.15)^{-3.06}.\label{EqNnu100}
\end{equation}
The dashed line in
Fig.~\ref{FigfV200} shows that the probability of hosting at most $3$
massive subhalos drops significantly when the $r_{100}$ radius
is used; only about $20\%$ of $M_{200}=10^{12}\,M_\odot$ halos pass
the test then. This stricter constraint emphasizes the difficulty of
resolving the missing massive satellite problem if the Milky Way mass
significantly exceeds $10^{12}\, M_{\odot}$.

\section{Summary}
\label{SecConc}

We have used the Millennium Simulation series, together with the
ultra-high resolution simulations of small halo samples from the
Aquarius and Phoenix projects to study the abundance of rare, massive
subhalos in $\Lambda$CDM halos. As in earlier work, we find that the
scaled subhalo velocity function (i.e., the number of subhalos as a
function of the ratio between subhalo maximum circular velocity and
host halo virial velocity, $\nu=V_{\rm max}/V_{200}$) is independent
of halo mass. This implies that we can obtain robust estimates of the
statistical distribution of massive subhalos from large samples of
well-resolved halos selected from the Millennium simulations.

Our main result is that, in the range $0.1<\nu<0.5$, the number of
subhalos within the virial radius, $r_{200}$, is
Poisson-distributed around an average given by
Eqn.~\ref{EqNnu}. Compared to this average, subhalos in the Aquarius
Project are slightly overabundant but still consistent given the large
variance and the small sample of halos included in that simulation
suite. Subhalos in the cluster-sized Phoenix Project halos are in
excellent agreement with Eqn.~\ref{EqNnu}.

We have then used this result to compute the probability that a halo
of virial velocity $V_{200}$ has a certain number of massive subhalos
with $V_{\rm max}$ exceeding a velocity threshold, $V_{\rm
  th}$. Applied to the Milky Way, where observations suggest that no
more than $3$ (or at most $4$) subhalos with $V_{\rm max}>30$ km/s
host luminous satellites, we find that this constraint effectively
translates into a strong upper limit on the Milky Way halo mass. The
probability that a halo with $M_{200} \gsim \, 3\times 10^{12} \,
M_\odot$ satisfies this constraint within radius $r_{200}$ is
vanishingly small, but it increases rapidly with decreasing virial
mass. Roughly $45\%$ of $M_{200}=10^{12}\, M_\odot$ halos pass this
test, and $\sim 90\%$ of all halos with $M_{200}\sim 5\times 10^{11}\,
M_\odot$ are consistent with the data. These fractions are reduced to
$\sim 20\%$ and $\sim 70\%$, respectively, if subhalos are considered
within a larger search radius, $r_{100}\sim 1.3 \, r_{200}$ (which,
for halos of mass $\sim 10^{12}\, M_\odot$, is close to the
Galactocentric distance of Leo I, the most distant bright satellite
known in the Milky Way). In this case, the number of subhalos,
$\langle N_{\rm sub}\rangle (>\nu)$, within $r_{100}$ is given by
Eqn.~\ref{EqNnu100} and a Milky Way halo mass of $M_{200}=2\times
10^{12}\, M_\odot$ is strongly ruled out by the satellite data.

The ``missing massive satellites problem'' in the Milky Way halo
highlighted by \citet{Boylan-Kolchin2011a,Boylan-Kolchin2011b} and by
\citet{Parry2011} may thus be resolved if the mass of the
Milky Way halo is $\sim 10^{12}\, M_\odot$ \citep[see
also][]{Vera-Ciro2012}. This is well within the range of halo masses
allowed by the latest estimates based on either the timing argument
\citep{Li2008} or on abundance-matching methods \citep{Guo2010}. It is
in even better agreement with the lower virial masses reported by
estimates based on (i) the radial velocity dispersion of Milky Way
satellites and halo stars \citep{Battaglia2005,Sales2007}; (ii) the
escape speed in the solar neighbourhood \citep{Smith2007}; or (iii)
the kinematics of halo blue horizontal branch stars \citep{Xue2008}.
Invoking a $\sim 10^{12}\, M_\odot$ mass for the Milky Way is a
simpler and more straightforward resolution than several alternatives
advanced in recent papers, such as considering the baryon adiabatic
contraction and feedback \citep{diCintio2011}, reducing the central density of
subhalos through tidal stripping \citep{diCintio2012,Vera-Ciro2012}, 
or positing radical revisions to the nature of dark matter
\citep{Lovell2012,Vogelsberger2012}.

We conclude that there is no compelling requirement to revise the
$\Lambda$CDM paradigm based on the abundance of massive subhalos in
the Milky Way. There are still, however, some uncomfortable
corollaries to this solution. One is that a $10^{12}\, M_\odot$ halo
has a virial velocity of only $\sim 150$ km/s, well below the rotation
speed of the Milky Way disk, usually assumed to be $V_{\rm rot}=220$
km/s, or even higher \citep{Reid2009}. This seems at odds with results
from some semi-analytic models of galaxy formation that attempt
simultaneously to match the Tully-Fisher relation and the galaxy
stellar mass function: agreement with observation seems to require
$V_{\rm rot}\approx V_{200}$ \citep[see,
e.g.,][]{Cole2000,Croton2006}.

A further worry is that an $M_{200}=10^{12} M_\odot$ halo might not be
massive enough to host satellites as massive as the Magellanic
Clouds. Assuming that $V_{\rm max}$ for the LMC and SMC can be
identified with the rotation speed of their HI disks, or $60$ and $50$
km/s, respectively \citep{Kim1998,Stanimirovic2004}, we find, using
the data in Table~\ref{TabN200Min}, that only $\sim 62\%$ of
$V_{200}=150$ km/s halos would be expected not to host an LMC-like
system. The probability of hosting two (or more) subhalos more massive
than the SMC is of order $20\%$. None of these probabilities seem
unlikely enough to cause worry.


Although our results may explain why few massive subhalos might be
expected in the Milky Way halo, this explanation still assigns MW
satellites to very low mass halos, i.e., those with $V_{\rm max}<30$
km/s. These halos have masses below $10^{10} M_\odot$, the mass scale
below which semi-analytic models predict that galaxy formation
efficiency should become exceedingly small \citep{Guo2010}.  Given the
large number of low mass halos expected in a $\Lambda$CDM universe,
populating even a small fraction of $V_{\rm max}<30$ km/s systems with
galaxies as bright as Fornax might lead to substantially
overpredicting the number of dwarfs in the local Universe \citep[see,
e.g.,][]{Ferrero2011}. Without a full accounting of how dwarf galaxies
form in low-mass halos, concerns about the viability of $\Lambda$CDM
on small scales will be hard to dispel.

\begin{table*}
  \caption{$N_{200}^{\rm  min}(\nu)$ is the minimum number of particles within the virial radius of a halo needed  to
    achieve convergence in the abundance of subhalos. $N_{\rm halos}$
    is the number of halos that satisfy such condition in the
    Millennium Simulations. $\langle N_{\rm sub} \rangle$ and $\sigma_{N_{\rm sub}}$ are the
    average number of subhalos exceeding $\nu$ and its dispersion, respectively.}
\begin{center}
  \begin{tabular}{cccccccccc} \hline
    $\nu$ & 0.1 & 0.15  & 0.2 & 0.25 & 0.3 & 0.35 & 0.4 & 0.45 & 0.5    \\
    $(V_{\rm max}/V_{200}$) & & & & & & & & &    \\ \hline \hline
   $N_{200}^{\rm min}$ &$1.0\times 10^6$ &  $2.5\times 10^5$ &  $1.2 \times
   10^5$ &  $7.5 \times 10^4$ &  $2.5 \times 10^4$ &  $1.8 \times 10^4$ &
   $1.0 \times 10^4$ &  $7.5 \times 10^3$ &  $5.0 \times 10^3$\\
    $N_{\rm halos}$&614 &     3070  &    6867  &   12138   &  38550  &
    54568  &   90200 &   113585&    151663\\ \hline
    $\langle N_{\rm sub} \rangle \, (r<r_{200})$ & 36.55 &    10.14 &     4.20 &     2.12
&    1.14 &     0.71 &     0.48 &     0.34  &    0.25\\
    $\sigma_{N_{\rm sub}}\, (r<r_{200})$& 8.92 & 3.87 & 2.27 &  1.54 &  1.10 &
    0.85  &   0.68 &   0.57  &    0.48\\ \hline
     $\langle N_{\rm sub} \rangle (r<r_{100})$ &51.91&15.22&6.3&3.11&1.73&1.06&0.69&0.46&0.33\\
    $\sigma_{N_{\rm sub}}\, (r<r_{100})$ &12.2&5.23&2.99&1.98&1.39&1.08&0.84&0.68&0.58\\
    \hline
  \end{tabular}
\end{center}
\label{TabN200Min}
\end{table*}

\section*{Acknowledgements}
We thank Mike Boylan-Kolchin, Volker Springel and Simon White for
useful suggestions and comments on early versions of this work. We
also thank an anonymous referee for helpful comments which helped
improve this paper. The
simulations analyzed in this paper were carried out by the Virgo
consortium for cosmological simulations and we are grateful to our
consortium colleagues for permission to use the data. Some of the
calculations were performed on the DiRAC facility jointly funded by
STFC, the Large Facilities Capital Fund of the department for
Business, Innovation and Skills and Durham University. JW
acknowledges a Royal Society Newton International Fellowship and CSF
a Royal Society Wolfson Research Merit Award This work was supported
by ERC Advanced Investigator grant COSMIWAY and an STFC rolling
grant to the Institute for Computational Cosmology. LG acknowledges
support from the one-hundred-talents program of the Chinese academy
of science (CAS), MPG partner Group family, the National Basic
Research Program of China (program 973 under grant No. 2009CB24901),
NSFC grants (Nos. 10973018 and 11133003) and an STFC Advanced
Fellowship, as well as the hospitality of the Institute for
Computational Cosmology at Durham University.


\bibliographystyle{mn2e}
\bibliography{master}

\begin{thebibliography}{}

\bibitem[\protect\citeauthoryear{{Battaglia}, {Helmi}, {Morrison}, {Harding},
  {Olszewski}, {Mateo}, {Freeman}, {Norris} \& {Shectman}}{{Battaglia}
  et~al.}{2005}]{Battaglia2005}
{Battaglia} G.,  {Helmi} A.,  {Morrison} H.,  {Harding} P.,  {Olszewski} E.~W.,
   {Mateo} M.,  {Freeman} K.~C.,  {Norris} J.,    {Shectman} S.~A.,  2005,
  \mnras, 364, 433

\bibitem[\protect\citeauthoryear{{Benson}, {Lacey}, {Baugh}, {Cole} \&
  {Frenk}}{{Benson} et~al.}{2002}]{Benson2002}
{Benson} A.~J.,  {Lacey} C.~G.,  {Baugh} C.~M.,  {Cole} S.,    {Frenk} C.~S.,
  2002, \mnras, 333, 156

\bibitem[\protect\citeauthoryear{{Boylan-Kolchin}, {Bullock} \&
  {Kaplinghat}}{{Boylan-Kolchin} et~al.}{2011}]{Boylan-Kolchin2011a}
{Boylan-Kolchin} M.,  {Bullock} J.~S.,    {Kaplinghat} M.,  2011, \mnras, 415,
  L40

\bibitem[\protect\citeauthoryear{{Boylan-Kolchin}, {Bullock} \&
  {Kaplinghat}}{{Boylan-Kolchin} et~al.}{2012}]{Boylan-Kolchin2011b}
{Boylan-Kolchin} M.,  {Bullock} J.~S.,    {Kaplinghat} M.,  2012, \mnras,
  p.~2657

\bibitem[\protect\citeauthoryear{{Boylan-Kolchin}, {Springel}, {White} \&
  {Jenkins}}{{Boylan-Kolchin} et~al.}{2010}]{Boylan-Kolchin2010}
{Boylan-Kolchin} M.,  {Springel} V.,  {White} S.~D.~M.,    {Jenkins} A.,  2010,
  \mnras, 406, 896

\bibitem[\protect\citeauthoryear{{Boylan-Kolchin}, {Springel}, {White},
  {Jenkins} \& {Lemson}}{{Boylan-Kolchin} et~al.}{2009}]{Boylan-Kolchin2009}
{Boylan-Kolchin} M.,  {Springel} V.,  {White} S.~D.~M.,  {Jenkins} A.,
  {Lemson} G.,  2009, \mnras, 398, 1150

\bibitem[\protect\citeauthoryear{{Bullock}, {Kravtsov} \& {Weinberg}}{{Bullock}
  et~al.}{2000}]{Bullock2000}
{Bullock} J.~S.,  {Kravtsov} A.~V.,    {Weinberg} D.~H.,  2000, \apj, 539, 517

\bibitem[\protect\citeauthoryear{{Cole}, {Lacey}, {Baugh} \& {Frenk}}{{Cole}
  et~al.}{2000}]{Cole2000}
{Cole} S.,  {Lacey} C.~G.,  {Baugh} C.~M.,    {Frenk} C.~S.,  2000, \mnras,
  319, 168

\bibitem[\protect\citeauthoryear{{Cooper}, {Cole}, {Frenk}, {White}, {Helly},
  {Benson}, {De Lucia}, {Helmi}, {Jenkins}, {Navarro}, {Springel} \&
  {Wang}}{{Cooper} et~al.}{2010}]{Cooper2010}
{Cooper} A.~P.,  {Cole} S.,  {Frenk} C.~S.,  {White} S.~D.~M.,  {Helly} J.,
  {Benson} A.~J.,  {De Lucia} G.,  {Helmi} A.,  {Jenkins} A.,  {Navarro} J.~F.,
   {Springel} V.,    {Wang} J.,  2010, \mnras, 406, 744

\bibitem[\protect\citeauthoryear{{Croton}, {Springel}, {White}, {De Lucia},
  {Frenk}, {Gao}, {Jenkins}, {Kauffmann}, {Navarro} \& {Yoshida}}{{Croton}
  et~al.}{2006}]{Croton2006}
{Croton} D.~J.,  {Springel} V.,  {White} S.~D.~M.,  {De Lucia} G.,  {Frenk}
  C.~S.,  {Gao} L.,  {Jenkins} A.,  {Kauffmann} G.,  {Navarro} J.~F.,
  {Yoshida} N.,  2006, \mnras, 365, 11

\bibitem[\protect\citeauthoryear{{Di Cintio}, {Knebe}, {Libeskind}, {Brook},
  {Yepes}, {Gottloeber} \& {Hoffman}}{{Di Cintio} et~al.}{2012}]{diCintio2012}
{Di Cintio} A.,  {Knebe} A.,  {Libeskind} N.~I.,  {Brook} C.,  {Yepes} G.,
  {Gottloeber} S.,    {Hoffman} Y.,  2012, ArXiv e-prints

\bibitem[\protect\citeauthoryear{{di Cintio}, {Knebe}, {Libeskind}, {Yepes},
  {Gottl{\"o}ber} \& {Hoffman}}{{di Cintio} et~al.}{2011}]{diCintio2011}
{di Cintio} A.,  {Knebe} A.,  {Libeskind} N.~I.,  {Yepes} G.,  {Gottl{\"o}ber}
  S.,    {Hoffman} Y.,  2011, \mnras, 417, L74

\bibitem[\protect\citeauthoryear{{Ferrero}, {Abadi}, {Navarro}, {Sales} \&
  {Gurovich}}{{Ferrero} et~al.}{2011}]{Ferrero2011}
{Ferrero} I.,  {Abadi} M.~G.,  {Navarro} J.~F.,  {Sales} L.~V.,    {Gurovich}
  S.,  2011, ArXiv e-prints

\bibitem[\protect\citeauthoryear{{Font}, {Benson}, {Bower}, {Frenk}, {Cooper},
  {De Lucia}, {Helly}, {Helmi}, {Li}, {McCarthy}, {Navarro}, {Springel},
  {Starkenburg}, {Wang} \& {White}}{{Font} et~al.}{2011}]{Font2011a}
{Font} A.~S.,  {Benson} A.~J.,  {Bower} R.~G.,  {Frenk} C.~S.,  {Cooper} A.,
  {De Lucia} G.,  {Helly} J.~C.,  {Helmi} A.,  {Li} Y.-S.,  {McCarthy} I.~G.,
  {Navarro} J.~F.,  {Springel} V.,  {Starkenburg} E.,  {Wang} J.,    {White}
  S.~D.~M.,  2011, \mnras, 417, 1260

\bibitem[\protect\citeauthoryear{{Gao}, {Navarro}, {Frenk}, {Jenkins},
  {Springel} \& {White}}{{Gao} et~al.}{2012}]{Gao2012}
{Gao} L.,  {Navarro} J.~F.,  {Frenk} C.~S.,  {Jenkins} A.,  {Springel} V.,
  {White} S.~D.~M.,  2012, ArXiv e-prints

\bibitem[\protect\citeauthoryear{{Guo}, {White}, {Boylan-Kolchin}, {De Lucia},
  {Kauffmann}, {Lemson}, {Li}, {Springel} \& {Weinmann}}{{Guo}
  et~al.}{2011}]{Guo2011a}
{Guo} Q.,  {White} S.,  {Boylan-Kolchin} M.,  {De Lucia} G.,  {Kauffmann} G.,
  {Lemson} G.,  {Li} C.,  {Springel} V.,    {Weinmann} S.,  2011, \mnras, 413,
  101

\bibitem[\protect\citeauthoryear{{Guo}, {White}, {Li} \&
  {Boylan-Kolchin}}{{Guo} et~al.}{2010}]{Guo2010}
{Guo} Q.,  {White} S.,  {Li} C.,    {Boylan-Kolchin} M.,  2010, \mnras, pp
  367--376

\bibitem[\protect\citeauthoryear{{Kauffmann}, {White} \&
  {Guiderdoni}}{{Kauffmann} et~al.}{1993}]{Kauffmann1993}
{Kauffmann} G.,  {White} S.~D.~M.,    {Guiderdoni} B.,  1993, \mnras, 264, 201

\bibitem[\protect\citeauthoryear{{Kim}, {Staveley-Smith}, {Dopita}, {Freeman},
  {Sault}, {Kesteven} \& {McConnell}}{{Kim} et~al.}{1998}]{Kim1998}
{Kim} S.,  {Staveley-Smith} L.,  {Dopita} M.~A.,  {Freeman} K.~C.,  {Sault}
  R.~J.,  {Kesteven} M.~J.,    {McConnell} D.,  1998, \apj, 503, 674

\bibitem[\protect\citeauthoryear{{Komatsu}, {Smith}, {Dunkley}, {Bennett},
  {Gold}, {Hinshaw}, {Jarosik}, {Larson}, {Nolta}, {Page}, {Spergel} \& et
  al.}{{Komatsu} et~al.}{2011}]{Komatsu2011}
{Komatsu} E.,  {Smith} K.~M.,  {Dunkley} J.,  {Bennett} C.~L.,  {Gold} B.,
  {Hinshaw} G.,  {Jarosik} N.,  {Larson} D.,  {Nolta} M.~R.,  {Page} L.,
  {Spergel} D.~N.,    et al. 2011, \apjs, 192, 18

\bibitem[\protect\citeauthoryear{{Kravtsov}, {Berlind}, {Wechsler}, {Klypin},
  {Gottl{\"o}ber}, {Allgood} \& {Primack}}{{Kravtsov}
  et~al.}{2004}]{Kravtsov2004}
{Kravtsov} A.~V.,  {Berlind} A.~A.,  {Wechsler} R.~H.,  {Klypin} A.~A.,
  {Gottl{\"o}ber} S.,  {Allgood} B.,    {Primack} J.~R.,  2004, \apj, 609, 35

\bibitem[\protect\citeauthoryear{{Li}, {De Lucia} \& {Helmi}}{{Li}
  et~al.}{2010}]{Li2010}
{Li} Y.-S.,  {De Lucia} G.,    {Helmi} A.,  2010, \mnras, 401, 2036

\bibitem[\protect\citeauthoryear{{Li} \& {White}}{{Li} \&
  {White}}{2008}]{Li2008}
{Li} Y.-S.,  {White} S.~D.~M.,  2008, \mnras, 384, 1459

\bibitem[\protect\citeauthoryear{{{\L}okas}}{{{\L}okas}}{2009}]{Lokas2009}
{{\L}okas} E.~L.,  2009, \mnras, 394, L102

\bibitem[\protect\citeauthoryear{{Lovell}, {Eke}, {Frenk}, {Gao}, {Jenkins},
  {Theuns}, {Wang}, {White}, {Boyarsky} \& {Ruchayskiy}}{{Lovell}
  et~al.}{2012}]{Lovell2012}
{Lovell} M.~R.,  {Eke} V.,  {Frenk} C.~S.,  {Gao} L.,  {Jenkins} A.,  {Theuns}
  T.,  {Wang} J.,  {White} S.~D.~M.,  {Boyarsky} A.,    {Ruchayskiy} O.,  2012,
  \mnras, 420, 2318

\bibitem[\protect\citeauthoryear{{Macci{\`o}}, {Kang}, {Fontanot},
  {Somerville}, {Koposov} \& {Monaco}}{{Macci{\`o}} et~al.}{2010}]{Maccio2010}
{Macci{\`o}} A.~V.,  {Kang} X.,  {Fontanot} F.,  {Somerville} R.~S.,  {Koposov}
  S.,    {Monaco} P.,  2010, \mnras, 402, 1995

\bibitem[\protect\citeauthoryear{{Moore}, {Ghigna}, {Governato}, {Lake},
  {Quinn}, {Stadel} \& {Tozzi}}{{Moore} et~al.}{1999}]{Moore1999b}
{Moore} B.,  {Ghigna} S.,  {Governato} F.,  {Lake} G.,  {Quinn} T.,  {Stadel}
  J.,    {Tozzi} P.,  1999, \apjl, 524, L19

\bibitem[\protect\citeauthoryear{{Parry}, {Eke}, {Frenk} \& {Okamoto}}{{Parry}
  et~al.}{2012}]{Parry2011}
{Parry} O.~H.,  {Eke} V.~R.,  {Frenk} C.~S.,    {Okamoto} T.,  2012, \mnras,
  419, 3304

\bibitem[\protect\citeauthoryear{{Pe{\~n}arrubia}, {McConnachie} \&
  {Navarro}}{{Pe{\~n}arrubia} et~al.}{2008}]{Penarrubia2008a}
{Pe{\~n}arrubia} J.,  {McConnachie} A.~W.,    {Navarro} J.~F.,  2008, \apj,
  672, 904

\bibitem[\protect\citeauthoryear{{Pe{\~n}arrubia}, {Navarro} \&
  {McConnachie}}{{Pe{\~n}arrubia} et~al.}{2008}]{Penarrubia2008b}
{Pe{\~n}arrubia} J.,  {Navarro} J.~F.,    {McConnachie} A.~W.,  2008, \apj,
  673, 226

\bibitem[\protect\citeauthoryear{{Reid}, {Menten}, {Zheng}, {Brunthaler},
  {Moscadelli}, {Xu}, {Zhang}, {Sato}, {Honma}, {Hirota}, {Hachisuka}, {Choi},
  {Moellenbrock} \& {Bartkiewicz}}{{Reid} et~al.}{2009}]{Reid2009}
{Reid} M.~J.,  {Menten} K.~M.,  {Zheng} X.~W.,  {Brunthaler} A.,  {Moscadelli}
  L.,  {Xu} Y.,  {Zhang} B.,  {Sato} M.,  {Honma} M.,  {Hirota} T.,
  {Hachisuka} K.,  {Choi} Y.~K.,  {Moellenbrock} G.~A.,    {Bartkiewicz} A.,
  2009, \apj, 700, 137

\bibitem[\protect\citeauthoryear{{Sales}, {Navarro}, {Abadi} \&
  {Steinmetz}}{{Sales} et~al.}{2007}]{Sales2007}
{Sales} L.~V.,  {Navarro} J.~F.,  {Abadi} M.~G.,    {Steinmetz} M.,  2007,
  \mnras, 379, 1464

\bibitem[\protect\citeauthoryear{{Smith}, {X.} \& {et al.}}{{Smith}
  et~al.}{2007}]{Smith2007}
{Smith} M.~C.,  {X.}   {et al.} 2007, \mnras, 379, 755

\bibitem[\protect\citeauthoryear{{Somerville}}{{Somerville}}{2002}]{Somerville%
2002}
{Somerville} R.~S.,  2002, \apjl, 572, L23

\bibitem[\protect\citeauthoryear{{Springel}, {Wang}, {Vogelsberger}, {Ludlow},
  {Jenkins}, {Helmi}, {Navarro}, {Frenk} \& {White}}{{Springel}
  et~al.}{2008}]{Springel2008b}
{Springel} V.,  {Wang} J.,  {Vogelsberger} M.,  {Ludlow} A.,  {Jenkins} A.,
  {Helmi} A.,  {Navarro} J.~F.,  {Frenk} C.~S.,    {White} S.~D.~M.,  2008,
  \mnras, 391, 1685

\bibitem[\protect\citeauthoryear{{Springel}, {White}, {Jenkins}, {Frenk},
  {Yoshida}, {Gao}, {Navarro}, {Thacker}, {Croton}, {Helly}, {Peacock}, {Cole},
  {Thomas}, {Couchman}, {Evrard}, {Colberg} \& {Pearce}}{{Springel}
  et~al.}{2005}]{Springel2005a}
{Springel} V.,  {White} S.~D.~M.,  {Jenkins} A.,  {Frenk} C.~S.,  {Yoshida} N.,
   {Gao} L.,  {Navarro} J.,  {Thacker} R.,  {Croton} D.,  {Helly} J.,
  {Peacock} J.~A.,  {Cole} S.,  {Thomas} P.,  {Couchman} H.,  {Evrard} A.,
  {Colberg} J.,    {Pearce} F.,  2005, \nat, 435, 629

\bibitem[\protect\citeauthoryear{{Springel}, {Yoshida} \& {White}}{{Springel}
  et~al.}{2001}]{Springel2001a}
{Springel} V.,  {Yoshida} N.,    {White} S.~D.~M.,  2001, New Astronomy, 6, 79

\bibitem[\protect\citeauthoryear{{Stanimirovi{\'c}}, {Staveley-Smith} \&
  {Jones}}{{Stanimirovi{\'c}} et~al.}{2004}]{Stanimirovic2004}
{Stanimirovi{\'c}} S.,  {Staveley-Smith} L.,    {Jones} P.~A.,  2004, \apj,
  604, 176

\bibitem[\protect\citeauthoryear{{Strigari}, {Bullock}, {Kaplinghat}, {Simon},
  {Geha}, {Willman} \& {Walker}}{{Strigari} et~al.}{2008}]{Strigari2008}
{Strigari} L.~E.,  {Bullock} J.~S.,  {Kaplinghat} M.,  {Simon} J.~D.,  {Geha}
  M.,  {Willman} B.,    {Walker} M.~G.,  2008, \nat, 454, 1096

\bibitem[\protect\citeauthoryear{{Strigari}, {Frenk} \& {White}}{{Strigari}
  et~al.}{2010}]{Strigari2010}
{Strigari} L.~E.,  {Frenk} C.~S.,    {White} S.~D.~M.,  2010, \mnras, 408, 2364

\bibitem[\protect\citeauthoryear{{Vera-Ciro}, {Helmi}, {Starkenburg} \&
  {Breddels}}{{Vera-Ciro} et~al.}{2012}]{Vera-Ciro2012}
{Vera-Ciro} C.~A.,  {Helmi} A.,  {Starkenburg} E.,    {Breddels} M.~A.,  2012,
  ArXiv e-prints

\bibitem[\protect\citeauthoryear{{Vogelsberger}, {Zavala} \&
  {Loeb}}{{Vogelsberger} et~al.}{2012}]{Vogelsberger2012}
{Vogelsberger} M.,  {Zavala} J.,    {Loeb} A.,  2012, ArXiv e-prints

\bibitem[\protect\citeauthoryear{{Walker}, {Mateo}, {Olszewski},
  {Pe{\~n}arrubia}, {Wyn Evans} \& {Gilmore}}{{Walker}
  et~al.}{2009}]{Walker2009}
{Walker} M.~G.,  {Mateo} M.,  {Olszewski} E.~W.,  {Pe{\~n}arrubia} J.,  {Wyn
  Evans} N.,    {Gilmore} G.,  2009, \apj, 704, 1274

\bibitem[\protect\citeauthoryear{{Weinberg}, {Colombi}, {Dav{\'e}} \&
  {Katz}}{{Weinberg} et~al.}{2008}]{Weinberg2008}
{Weinberg} D.~H.,  {Colombi} S.,  {Dav{\'e}} R.,    {Katz} N.,  2008, \apj,
  678, 6

\bibitem[\protect\citeauthoryear{{White} \& {Frenk}}{{White} \&
  {Frenk}}{1991}]{white1991}
{White} S.~D.~M.,  {Frenk} C.~S.,  1991, \apj, 379, 52

\bibitem[\protect\citeauthoryear{{Wolf}, {Martinez}, {Bullock}, {Kaplinghat},
  {Geha}, {Mu{\~n}oz}, {Simon} \& {Avedo}}{{Wolf} et~al.}{2010}]{Wolf2010}
{Wolf} J.,  {Martinez} G.~D.,  {Bullock} J.~S.,  {Kaplinghat} M.,  {Geha} M.,
  {Mu{\~n}oz} R.~R.,  {Simon} J.~D.,    {Avedo} F.~F.,  2010, \mnras, 406, 1220

\bibitem[\protect\citeauthoryear{{Xue}, {Rix}, {Zhao}, {Re Fiorentin}, {Naab},
  {Steinmetz}, {van den Bosch}, {Beers}, {Lee}, {Bell}, {Rockosi}, {Yanny},
  {Newberg}, {Wilhelm}, {Kang}, {Smith} \& {Schneider}}{{Xue}
  et~al.}{2008}]{Xue2008}
{Xue} X.~X.,  {Rix} H.~W.,  {Zhao} G.,  {Re Fiorentin} P.,  {Naab} T.,
  {Steinmetz} M.,  {van den Bosch} F.~C.,  {Beers} T.~C.,  {Lee} Y.~S.,  {Bell}
  E.~F.,  {Rockosi} C.,  {Yanny} B.,  {Newberg} H.,  {Wilhelm} R.,  {Kang} X.,
  {Smith} M.~C.,    {Schneider} D.~P.,  2008, \apj, 684, 1143

\bibitem[\protect\citeauthoryear{{Zheng}, {Berlind}, {Weinberg}, {Benson},
  {Baugh}, {Cole}, {Dav{\'e}}, {Frenk}, {Katz} \& {Lacey}}{{Zheng}
  et~al.}{2005}]{Zheng2005}
{Zheng} Z.,  {Berlind} A.~A.,  {Weinberg} D.~H.,  {Benson} A.~J.,  {Baugh}
  C.~M.,  {Cole} S.,  {Dav{\'e}} R.,  {Frenk} C.~S.,  {Katz} N.,    {Lacey}
  C.~G.,  2005, \apj, 633, 791

\end{thebibliography}

\bsp

\label{lastpage}

\end{document}